\documentclass[%
 reprint,
nofootinbib,
 amsmath,amssymb,
 aps,
superscriptaddress]{revtex4-2}

\usepackage{graphicx,subfigure}
\usepackage{dcolumn,tabularx,calc,multirow,booktabs,slashed}
\usepackage{bm}
\usepackage[colorlinks=true,citecolor=blue,urlcolor=cyan,filecolor=magenta,linkcolor=red,frenchlinks=true,bookmarks]{hyperref}

\begin{document}

\preprint{APS/123-QED}

\title{Electromagnetic form factors of $\Omega^-$ with the meson cloud in the spacelike and timelike regions}

\author{Dongyan Fu}

\affiliation{Southern Center for Nuclear-Science Theory (SCNT), Institute of Modern Physics, Chinese Academy of Sciences, Huizhou 516000, China}

\author{Ju-Jun Xie}~\email{xiejujun@impcas.ac.cn}

\affiliation{Southern Center for Nuclear-Science Theory (SCNT), Institute of Modern Physics, Chinese Academy of Sciences, Huizhou 516000, China}
\affiliation{Heavy Ion Science and Technology Key Laboratory, Institute of Modern Physics, Chinese Academy of Sciences, Lanzhou 730000, China}
\affiliation{School of Nuclear Sciences and Technology, University of Chinese Academy of Sciences, Beijing 101408, China}

\author{Yubing Dong}~\email{dongyb@ihep.ac.cn}

\affiliation{Institute of High Energy Physics, Chinese Academy of Sciences, Beijing 100049, China}
\affiliation{School of Physical Sciences, University of Chinese Academy of Sciences, Beijing 101408, China}

\date{\today}

\begin{abstract}
We present comprehensive calculations of the electromagnetic form factors, including the charge, magnetic-dipole, electric-quadrupole and magnetic-octupole form factors, of the $\Omega^-$ in the spacelike region using the quark-diquark approach, incorporating the effect of the meson cloud. The obtained magnetic moment is in agreement with the experimental measurement and the electromagnetic radii are comparable with other model calculations. It is found that the meson cloud effect remains almost unchanged as the energy becomes high and this feature implies that the meson cloud should be considered in all the energy region. Moreover, the asymptotic relations allow us to extend the electromagnetic form factors from the spacelike region to the timelike region, and then the effective form factor is almost identical with the data from CLEO and BESIII.
Finally, polarization properties of the final state $\Omega^-$ in the $e^+ e^- \rightarrow \Omega^- \bar{\Omega}^+$ process with the unpolarized initial states are also investigated.
\end{abstract}

\maketitle

\section{Introduction}

The electromagnetic structure of hyperons constitutes a fundamental frontier in hadronic physics, providing critical insights into the non-perturbative regime of quantum chromodynamics (QCD)~\cite{Brodsky:1980zm,Roberts:1994dr}. As key descriptors of the composite particle structure, electromagnetic form factors (EMFFs) have been instrumental in advancing our understanding of baryonic systems~\cite{Beg:1964nm,Capstick:2000qj}.
While octet baryon EMFFs have been thoroughly investigated through theoretical frameworks \cite{Isgur:1978xj,Wang:2024abv} and experimental measurements~\cite{JeffersonLabHallA:1999epl,Perdrisat:2006hj,Arrington:2007ux,BESIII:2021rqk}, the spin-3/2 decuplet particles such as $\Omega^-$ present unique challenges due to their enhanced tensor complexity \cite{Nozawa:1990gt,Butler:1993ej}. At leading order, the electromagnetic property of a spin-3/2 particle is characterized by four independent EMFFs, charge, electro-quadrupole, magnetic-dipole, and magnetic-octupole form factors. In the one photon exchange approximation, according to the four-momentum squared $q^2$ of the photon, these electromagnetic form factors are conventionally analyzed in two distinct kinematic domains: the spacelike region ($q^2 < 0$) accessible through electron hadron scattering, and the timelike region ($q^2 > 0$) probed via annihilation processes. Despite the first proton EMFF measurements dating back to 1951, experimental determination of spin-3/2 baryon EMFFs has been severely constrained by their short lifetimes. The $\Omega^-$ hyperon emerges as an exceptional candidate for such studies due to its relatively long lifetime ($c\tau = 2.461$ cm). However, the measurements of the $\Omega^-$ EMFFs in the spacelike region by the $e^- \Omega^-$ scattering is scarce since making $\Omega^-$ baryon as target is very difficult~\cite{Aznauryan:2011qj,Afanasev:2012fh}. Recent experimental breakthroughs using $e^+e^- \rightarrow \Omega^-\bar{\Omega}^+$ process have enabled pioneering timelike measurements. For examples, Ref.~\cite{Dobbs:2014ifa} firstly reported the effective form factors of $\Omega^-$ baryon at large timelike momentum transfers using data from CLEO, with subsequent precision measurements achieved by BESIII collaboration at different energy scales~\cite{BESIII:2022kzc,BESIII:2025azv}.

On the theoretical side, the schemes to calculate the EMFFs of $\Omega^-$ baryon include the lattice QCD simulations~\cite{Alexandrou:2010jv} and some theoretical models. The frequently-used models contain the chiral constituent quark model ($\chi$CQM)~\cite{Wagner:2000ii,Broniowski:2008hx,Kim:2020nug}, the chiral perturbation theory ($\chi$PT)~\cite{Butler:1993ej,Li:2016ezv}, the $1/N_c$ expansion~\cite{Luty:1994ub,Buchmann:2018nmu}, the SU(2) Skyme model~\cite{Kim:2012ts}, the bag model~\cite{Neubelt:2019sou}, the QCD sum rule (QCDSR)~\cite{Lee:1997jk,Azizi:2019ytx},  the relativistic quark model (RQM)~\cite{Schlumpf:1993rm,Ramalho:2009gk}, and the non-relativistic quark model (NRQM) \cite{Isgur:1981yz,Krivoruchenko:1991pm}. On the experimental side, the measurement of the $\Omega^-$ EMFFs in the timelike region can be done from the $e^+ e^- \to \Omega^- \bar{\Omega}^+$ reaction which is easier than the spacelike region. There are only a few successful theoretical model calculations of the baryon EMFFs in the timelike region, such as the final-state interaction approach~\cite{Haidenbauer:2016won,Dai:2024lau} and the vector-meson dominance model~\cite{Gounaris:1968mw,Yang:2019mzq,Chen:2023oqs,Yan:2023nlb,Chen:2024luh}.

In this work, following the previous works~\cite{Fu:2022rkn,Fu:2023ijy,Wang:2023bjp}, we aim to study the EMFFs of $\Omega^-$ baryon in both the spacelike and timelike regions using the quark-diquark approach and providing predictions for related observables. It is worthy to mention that the direct calculation of the EMFFs in timelike region using the quark-diquark approach is unfeasible, and the relation between the spacelike and timelike regions is needed. While the dispersion relation~\cite{Khodjamirian:2020btr} facilitate transformations from timelike to spacelike regions, the inverse mapping requires other alternative approaches. We finally adopt a novel strategy that connects spacelike results to the timelike regime through asymptotic relations~\cite{Ramalho:2020laj} which focuses on the perturbative regime of QCD. The improvement here compared to the previous works is that the internal structure of the constituent quark is also investigated using the dressed quark with the meson cloud~\cite{Cloet:2014rja} instead of the point quark. And we will take Kaon as the meson cloud because Kaon is the lightest meson containing strange quark. The dressed quark is expected to provide a more accurate description on the $\Omega^-$, then the dressed quark has size and provides and extra $i \sigma^{\mu q}/(2 M)$ ($M$ stands for the mass of $\Omega^-$) term in the electromagnetic interaction, which is not exist in the point-like quark and important to the physical quantities related with the magnetic form factors.

The present article is organized as follows. In Sec.~\ref{section2}, the EMFFs with the unified form in the spacelike and timelike regions and the effective form factor from extension of the spacelike region using the asymptotic relations are given. Moreover, we encapsulate the brief calculation process of the quark-diquark approach and elaborate the electromagnetic interaction vertex with the dressed quark with the meson cloud. In Sec.~\ref{section3}, by fitting the lattice EMFFs with different lattices, the parameters are determined and used when we extend the results to those under the physical masses. The EMFFs in the spacelike and timelike regions then are respectively shown and discussed. Finally, a brief summary and conclusion is presented in Sec.~\ref{section4}.

\section{Electromagnetic form factors of Spin-3/2 particle and quark-diquark approach} \label{section2}

\subsection{Electromagnetic structure of the spin-3/2 particle}

In the one photon approximation, a particle with spin-3/2 has four independent EMFFs, which are defined by the following matrix elements of the electromagnetic current in the spacelike (SL) and timelike (TL) regions,
\begin{widetext}
\begin{subequations}\label{Definiton_EMFFs}
	\begin{eqnarray}
 \left\langle N_{3/2}(p',\lambda') \left|J^\mu(0)\right|N_{3/2}(p,\lambda)\right\rangle =
 - \bar{u}_{\alpha'}(p',\lambda') \Gamma^{SL,\alpha' \alpha \mu} u_\alpha(p,\lambda),
	\end{eqnarray}
	and
	\begin{equation}
		\left\langle N_{3/2}(p',\lambda') \bar{N}_{3/2}(p,\lambda)\left|J^\mu(0)\right|0\right\rangle = - \bar{u}_{\alpha'}(p',\lambda') \Gamma^{TL,\alpha' \alpha \mu} v_\alpha(p,\lambda).
	\end{equation}
\end{subequations}
\end{widetext}
where $u_\alpha(p,\lambda)$ is Rarita-Schwinger spinor for a spin-3/2 particle, with the normalization $\bar{u}_{\alpha'}(p)u_\alpha(p) = - 2 M \delta_{\alpha' \alpha}$, with $M$ the mass of the baryon. The electromagnetic structures defined by Eq.~\eqref{Definiton_EMFFs} have the similar form,
\begin{subequations}\label{EM_Structure}
	\begin{eqnarray}
		\Gamma^{SL,\alpha' \alpha \mu} &=& \gamma^\mu \left(g^{\alpha' \alpha} F^{SL}_{1,0} +  \frac{q^{\alpha'}q^\alpha}{2 M^2}F^{SL}_{1,1}\right) + \nonumber \\
  &&  \frac{i \sigma^{\mu q}}{2 M} \left(g^{\alpha' \alpha} F^{SL}_{2,0} + \frac{q^{\alpha'}q^\alpha}{2 M^2}F^{SL}_{2,1}\right),
	\end{eqnarray}
	and
	\begin{eqnarray}
		\Gamma^{TL,\alpha' \alpha \mu} &=& \gamma^\mu \left(g^{\alpha' \alpha} F^{TL}_{1,0} + \frac{q^{\alpha'}q^\alpha}{2 M^2}F^{TL}_{1,1}\right) +  \nonumber \\
  && \frac{i \sigma^{\mu q}}{2 M} \left(g^{\alpha' \alpha} F^{TL}_{2,0} + \frac{q^{\alpha'}q^\alpha}{2 M^2}F^{TL}_{2,1}\right),
	\end{eqnarray}
\end{subequations}
where $q=p'-p$ ($q^2=-Q^2<0$) in the spacelike region and $q=p'+p$ ($q^2=-Q^2>0$) in the timelike region. 

Then, the obtained EMFFs with the electromagnetic structures shown in Eqs.~\eqref{EM_Structure} are as following~\cite{Korner:1976hv,Nozawa:1990gt},
\begin{eqnarray}
G_{E0}^{SL/TL} &=& \left(1-\frac{2}{3}\tau\right)\left(F_{1,0}^{SL/TL}+\tau F_{2,0}^{SL/TL}\right)+ \nonumber \\
  && \frac{2}{3}\tau(1-\tau)\left(F_{1,1}^{SL/TL}+\tau F_{2,1}^{SL/TL}\right), \\	
G_{E2}^{SL/TL} &=& \left(F_{1,0}^{SL/TL}+\tau F_{2,0}^{SL/TL}\right) -  \nonumber \\
  &&  (1-\tau)\left(F_{1,1}^{SL/TL}+\tau F_{2,1}^{SL/TL}\right), \\
G_{M1}^{SL/TL} &=& \left(1-\frac{4}{5}\tau\right)\left(F_{1,0}^{SL/TL}+F_{2,0}^{SL/TL}\right)+  \nonumber \\
  &&  \frac{4}{5}\tau(1-\tau)\left(F_{1,1}^{SL/TL}+F_{2,1}^{SL/TL}\right), \\
G_{M3}^{SL/TL} &=& \left(F_{1,0}^{SL/TL}+F_{2,0}^{SL/TL}\right)-  \nonumber \\
  &&  (1-\tau)\left(F_{1,1}^{SL/TL}+F_{2,1}^{SL/TL}\right),
\end{eqnarray}
where $\tau$ has the same form $\tau = q^2/(4 M^2)$ in both the spacelike and timelike regions~\footnote{For the sake of convenience in the presentation, we adopt the same notation $q$ for both the spacelike and timelike energy regions.}. The $G_{E0}$, $G_{E2}$, $G_{M1}$, and $G_{M3}$ are the charge, electric-quadrupole, magnetic-dipole and magnetic-octupole form factors, respectively. In the forward limit in the spacelike region, the momentum-transfer squared goes to zero, one can obtain the charge, electric-quadrupole, magnetic-dipole and magnetic-octupole moments.

On the other hand, we use the scheme in which the EMFFs in the spacelike region could be extended to the timelike region according to the asymptotic relations~\cite{Pacetti:2014jai,Ramalho:2020laj},
\begin{subequations}\label{SL_to_TL}
	\begin{equation}
		G_{E(0,2)}^{TL}(q^2) = G_{E(0,2)}^{SL}(-q^2 + 2 M^2),
	\end{equation}
	\begin{equation}
		G_{M(1,3)}^{TL}(q^2) = G_{M(1,3)}^{SL}(-q^2 + 2 M^2),
	\end{equation}
\end{subequations}
with $q^2 > 4 M^2$. The approximation relations in Eq.~\eqref{SL_to_TL} are derived from the unitarity of the electromagnetic current and the Phragmén-Lindelöf theorem, which implies that the limits for large \( |q^2| \) are identical for both the spacelike and timelike form factors~\cite{Denig:2012by,Pacetti:2014jai}. Meanwhile, the relations in Eq.~\eqref{SL_to_TL} entail the real timelike form factors and conceal the relevant phase information.

In the time-like region, for the $e^+ e^- \rightarrow \Omega^- \bar{\Omega}^+$ process with the condition that the initial $e^-$ and $e^+$ are unpolarized, the cross sections are obtained as~\cite{Korner:1976hv,BaBar:2005pon,Pacetti:2014jai,Dobbs:2017hyd,Ramalho:2020laj},
\begin{eqnarray}
		\frac{d \sigma^{un}}{d cos \theta} &=& \frac{\pi \alpha^2 \beta C}{2 q^2} \left[ \frac{1}{\tau} \left(2|G^{TL}_{E0}|^2 + \frac{8}{9} \tau^2 |G^{TL}_{E2}|^2\right)\text{sin}^2\theta + \right.\nonumber \\
  &&\left. \left(\frac{10}{9} |G^{TL}_{M1}|^2 + \frac{16}{15} \tau^2 |G^{TL}_{M3}|^2\right) (1+\text{cos}^2\theta) \right],
\end{eqnarray}
\begin{equation}
	\sigma(q^2)=\frac{4 \pi \alpha^2 \beta C}{3 q^2} \left(1+\frac{1}{2 \tau}\right)\left|G^{eff}_{EM}(q^2)\right|^2,
\end{equation}
where the polar angle $\theta$ is defined as the angle between the direction of the final $\Omega^-$ and the initial positron beam. The $\left|G^{eff}_{EM}(q^2)\right|$ is the effective form factor, $\alpha$ is the fine-structure constant, $\beta = \sqrt{1-1/\tau}$, $C=y/(1-e^{-y})$, and $y= \frac{\pi \alpha}{2 \sqrt{\tau - 1}}$, which is the Coulomb correction factor~\cite{Tzara:1970ne,BaBar:2005pon}. The effective form factor $\left|G^{eff}_{EM}(q^2)\right|$ is~\cite{Korner:1976hv,Ramalho:2020laj},
\begin{eqnarray} \label{Effective_EMFFs}
\left|G^{eff}_{EM}(q^2)\right|^2 &=& \left(1+\frac{1}{2 \tau}\right)^{-1} \left[\left(\frac{10}{9} |G^{TL}_{M1}|^2 + \frac{16}{15} \tau^2 |G^{TL}_{M3}|^2\right) \right. \nonumber \\
&& \left. + \frac{1}{2 \tau} \left(2|G^{TL}_{E0}|^2 + \frac{8}{9} \tau^2 |G^{TL}_{E2}|^2\right)\right].
\end{eqnarray}
Due to the different definition of the magnetic-octupole form factor $G^{TL}_{M3}$ from Refs.~\cite{Korner:1976hv,Ramalho:2020laj}, we take the coefficient as $\frac{16}{15}$ instead of $\frac{32}{5}$. Furthermore, the magnetic-octupole moments in Ref.~\cite{Korner:1976hv} and Ref.~\cite{Nozawa:1990gt} have a different factor $\sqrt{6}$, namely $G_{M3}$ in Ref.~\cite{Korner:1976hv} is $G_{M3}/\sqrt{6}$ of Ref.~\cite{Nozawa:1990gt}. The form of $G_{M3}$ in Ref.~\cite{Nozawa:1990gt} was commonly used in this work and these previous works~\cite{Fu:2022rkn,Fu:2023ijy,Wang:2023bjp}. On the other hand, Eq.~\eqref{Effective_EMFFs} imply that the highest order magnetic-octupole form factor $G_{M3}$ dominates the effective form factor as the energy increases. This character is consistent with the conclusion obtained in Ref.~\cite{Ramalho:2020laj}.

\subsection{Quark-siquark approach}

\begin{figure}[htbp]
    \begin{center}
        \subfigure[]{\includegraphics[width=7cm]{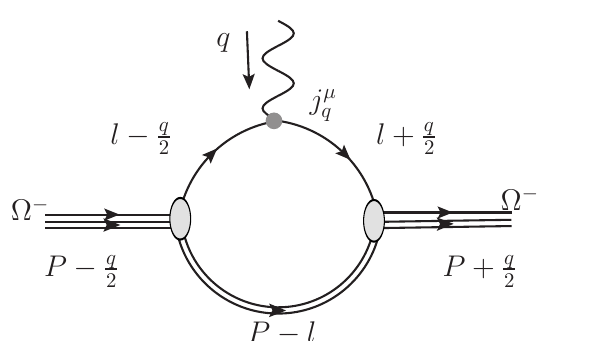}}
        \subfigure[]{\includegraphics[width=7cm]{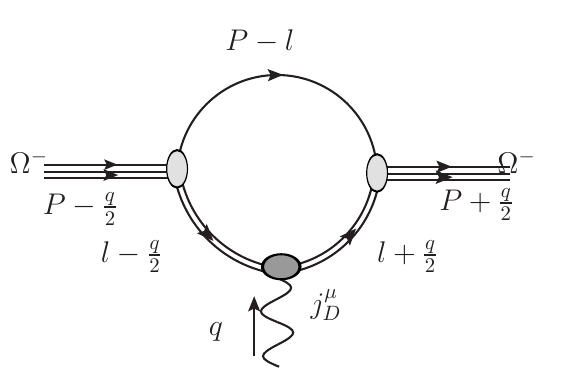}}
        \subfigure[]{\includegraphics[width=7cm]{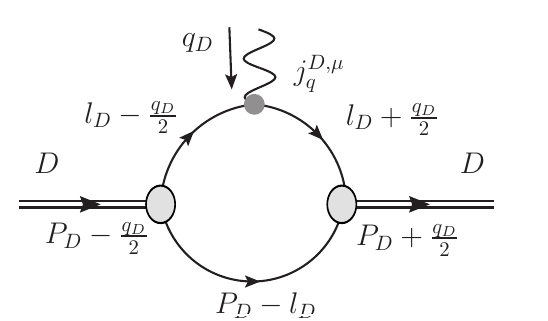}}
        \caption{\label{FeynmanOmegaEM}\small{Feynman diagrams for the electromagnetic current of the $\Omega^-$, (a) and (b), and of the diquark (c). The left and middle panels stand for the contributions of quark (single line) and diquark (double line) to $\Omega^-$.}}
    \end{center}
\end{figure}

The $\Omega^-$ hyperon with the quantum number $I(J^P)=0(\frac{3}{2}^+)$ is composed of three strange quarks. It is convenient to regard it as a bound state with one strange quark and one diquark, which consists of two strange quarks and has $J^P=1^+$. In addition, the internal structure of the axial-vector diquark also needs to be considered. The EMFFs of $\Omega^-$ baryon can be obtained from the matrix element of the electromagnetic current, 
which are illustrated by the Feynman diagrams in Fig.~\ref{FeynmanOmegaEM}. Then one can write down the electromagnetic current as
\begin{equation}
    \begin{split}
        & \left\langle p^\prime,\lambda^\prime \left| \hat{J}^{\mu}(0) \right| p,\lambda\right\rangle \\
        & = \left\langle p^\prime,\lambda^\prime
        \left| \hat{J}^{\mu}_{q}(0) \right| p,\lambda\right\rangle + \left\langle p^\prime,\lambda^\prime \left| \hat{J}^{\mu}_D(0)
        \right| p,\lambda\right\rangle,
    \end{split}
\end{equation}
where the matrix element is expressed as the sum of the quark and diquark contributions which are labeled by the subscripts, $q$ and $D$, respectively.

According to the Feynman diagram of Fig.~\ref{FeynmanOmegaEM} (a), the specific expression of the quark contribution can be obtained as
\begin{equation}\label{Omegaquarkj}
    \begin{split}
    & \left\langle p^\prime,\lambda^\prime \left| \hat{J}^{\mu}_{q}(0)
    \right| p,\lambda\right\rangle =  i c^2 eQ^e_{q}   \\ 
     & \times \bar{u}_{\alpha'}(p',\lambda')  \int \frac{d^4 l}{(2 \pi)^4}\frac{1}{\mathfrak{D}}
    \Gamma^{\alpha' \beta'}  g_{\beta' \beta}  \\
    & \times \left( \slashed{l}+\frac{\slashed{q}}{2}+m_q \right)
   j_q^\mu \left( \slashed{l}-\frac{\slashed{q}}{2}
    +m_q \right)\Gamma^{\alpha \beta} u_{\alpha}(p,\lambda),
    \end{split}
\end{equation}
where $Q_q^e$ is the electric charge number carried by the quark participating in the interaction, and the parameter $c$ is fixed in order to normalize the electric charge number of $\Omega^-$ at $Q^2=0$. $j_q^\mu$ is the interaction vertex between the photon field and the quark field and $j_q^\mu = \gamma^\mu$ for the point quark. In addition, $\mathfrak{D}$ and $\Gamma^{\alpha \beta}$ are given by
\begin{equation}\label{OmegaD}
    \begin{split}
    \mathfrak{D} & = [\left( l-P \right) ^2 -m_R^2+i \epsilon]^2
    [\left(l-P\right)^2-m_D^2+i \epsilon]\\
    & \times \biggl[\left( l-\frac{q}{2} \right) ^2 -m_R^2+i \epsilon\biggr]
    \biggl[\left( l+\frac{q}{2} \right) ^2 -m_R^2+i \epsilon\biggr] \\
    & \times \biggl[\left(l+ \frac{q}{2} \right)^2 - m_q^2+i \epsilon\biggr]
    \biggl[\left(l- \frac{q}{2} \right)^2 - m_q^2+i \epsilon\biggr].
    \end{split}
\end{equation}
and
\begin{equation} \label{vertexfunction}
    \Gamma^{\alpha \beta} = g^{\alpha \beta} + c_2 \gamma^\beta \Lambda^\alpha
    + c_3 \Lambda^\beta \Lambda^\alpha,
\end{equation}
where $\Lambda$ is the relative momentum between the quark and diquark, and the superscript $\alpha$ ($\beta$) represents the index of the $\Omega^-$ (diquark). $m_q$ and $m_D$ are the masses of the quark and diquark, respectively. The couplings, $c_2$ and $c_3$, can be determined by fitting them to the lattice QCD results for the EMFFs of $\Omega^-$ baryon. 

In addition, to avoid the loop integral divergence, we employ one simple regularization, i.e. add a scalar function,
\begin{equation}\label{vertexfunction2}
    \Xi (p_1,p_2)=\frac{c}{[ p_1^2 -m_R^2+i \epsilon][ p_2^2 -m_R^2+i \epsilon]},
\end{equation}
at each vertex, where $m_R$ is a cutoff mass parameter. It should be mentioned that this simplification may break the gauge invariance slightly, however it is simpler than other sophisticated methods, such as the Pauli-Villars regularization~\cite{Pauli:1949zm}.

\subsection{Kaon meson cloud}

\begin{figure}[htbp]
	\centering
	\includegraphics[width=7cm]{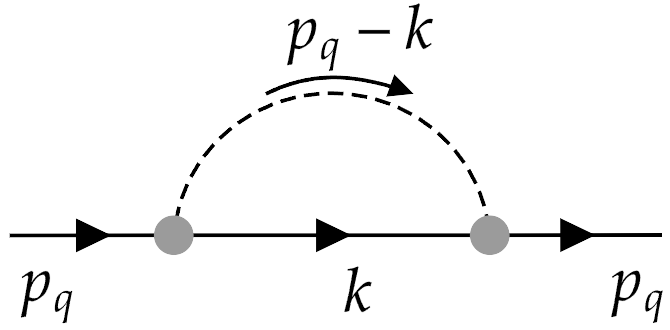}
	\caption{\label{figure-self-energy}\small{Kaon loop contribution to the dressed quark self-energy. The Kaon couples to the dressed quark via $\gamma_5$ and a pole form is employed as the Kaon t-matrix.}}
\end{figure}

In this work, we will also consider the effects of the meson cloud in the quark electromagnetic current. In other words, we will treat the quark as the dressed quark with the internal structure instead of the point particle. Because of the isospin conservation the strange quark cannot be dressed by the pion cloud and the Kaon meson cloud is the permissible meson with the lowest mass to dress the strange quark. Therefore, the Kaon meson cloud around the point quark is the principal part and we will only include the Kaon meson cloud in our calculation. Then, instead of the point form $\gamma^\mu$, we take the interaction vertex $j_q^\mu$ in Eq.~\eqref{Omegaquarkj} as the dressed strange quark form,
\begin{equation}\label{dressed-quark-vertex}
	j^\mu_s(p'_q,p_q)=\gamma^\mu F_{1 s}(q^2) + \frac{i \sigma^{\mu q}}{2 m_q} F_{2 s}(q^2),
\end{equation}
with $q=p'_q-p_q$. The first part of $F_{1 s}(q^2)$ is contributed from the point quark without the Kaon cloud. The probability to strike a dressed quark without the Kaon cloud can be obtained from the self-energy diagram of the dressed quark shown in Fig.~\ref{figure-self-energy}.

According to the self-energy diagram shown in Fig.~\ref{figure-self-energy}, the probability has the specific form,
\begin{equation}\label{self-energy-Z}
	Z = 1 + \left. \frac{\partial \Sigma(p_q)}{\partial{\slashed{p}_q}} \right|_{\slashed{p}_q=m_q},
\end{equation}
and the self-energy reads
\begin{equation}\label{self-energy}
	\Sigma(p_q) = -2 \int\frac{d^4 k}{(2 \pi)^4} \gamma_5 S(p_q-k) \gamma_5 \tau_K,
\end{equation}
with $S(k) = 1/(\slashed{k} -m_q +i \epsilon)$ the usual quark propagator with mass $m_q$. The factor $2$ is derived from the kinds of Kaon components, $\bar{u}s$ and $\bar{d}s$, containing $s$ quark. Moreover, the reduced Kaon t-matrix is approximated by its pole form
\begin{equation}\label{tauk}
	\tau_K \rightarrow \frac{i Z_K}{k^2-m_K^2 + i \epsilon}.
\end{equation}
The pole residue $Z_K$ is given by the bubble diagram of Kaon,
\begin{equation}\label{zk}
	Z_K ^{-1} = - \left. \frac{\partial}{\partial q^2} \Pi_{PP}(q^2)\right|_{q^2 = m_K^2},
\end{equation}
where
\begin{equation}
	\Pi_{PP}(q^2) = 6i \int \frac{d^4 k}{(2 \pi)^4} Tr \left[\gamma^5 S(k) \gamma^5 S(k+q) \right].
\end{equation}

\begin{figure}[h]
	\centering
\includegraphics[width=7cm]{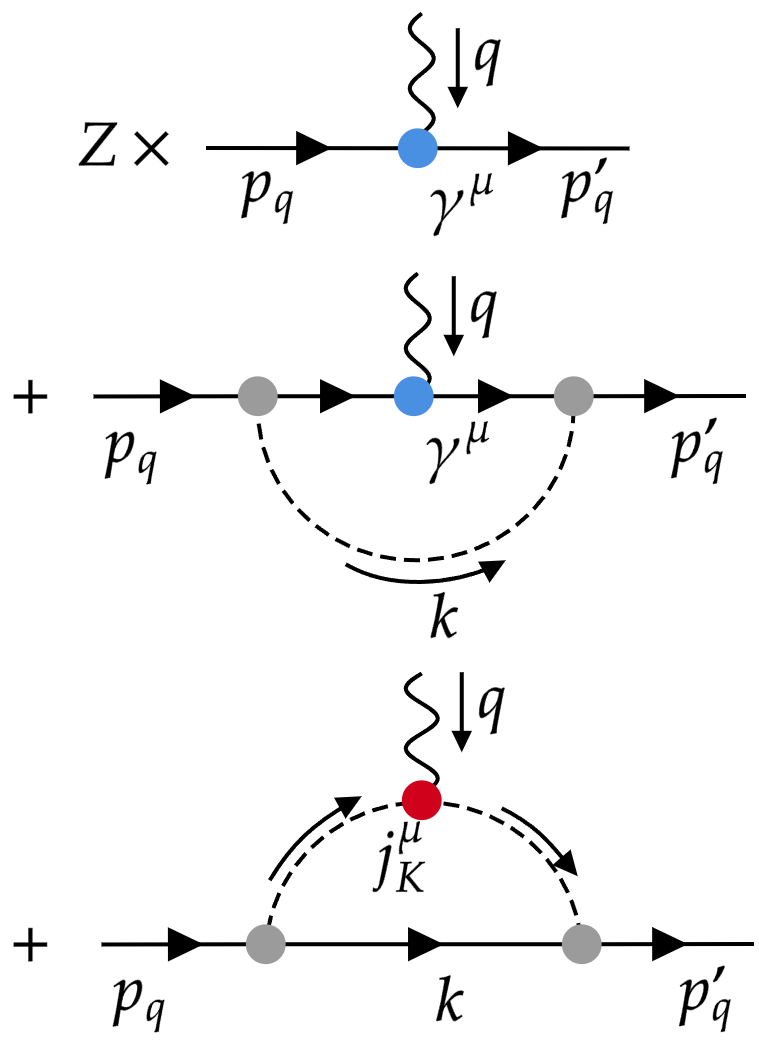}
\caption{\label{figure-kaon-loop}\small{Kaon loop contribution to the quark electromagnetic current. The quark wave function renormalization factor $Z$ represents the probability of striking a dressed quark without a kaon cloud. The blue shaded circle represents the electromagnetic vertex of the point quark and the red shaded circle the Kaon cloud. And the Kaon couples to the dressed quark via $\gamma_5$ marked as the gray shaded circle.}}
\end{figure}

Finally, the dressed quark electromagnetic current is illustrated in Fig.~\ref{figure-kaon-loop}. Then the $F_{1 s}(q^2)$ and $F_{2 s}(q^2)$ in Eq.~\eqref{dressed-quark-vertex} can be written as,
\begin{eqnarray}\label{dressed-F1F2}
	F_{1 s}(q^2) &=& -\frac{1}{3} Z + f_1^{(q)} + f_1^{(K)}, \\
        F_{2 s}(q^2) &=& f_2^{(q)} + f_2^{(K)},
\end{eqnarray}
where $q$ in the superscript represents $u$ or $d$ quark and $K$ is the corresponding Kaon. Notice that $f_{1,2}^{(q,K)}$ in Eq.~\eqref{dressed-F1F2} contains the probability to strike a dressed quark with the Kaon cloud.

In the dressed quark vertex part, the proper-time regularization scheme~\cite{Ebert:1996vx,Hellstern:1997nv,Bentz:2001vc,Cloet:2014rja} was chosen as
\begin{equation}\label{proper-time}
    \begin{split}
        \frac{1}{X^n} & = \frac{1}{(n-1)!}\int^\infty_0 d \tau \tau^{n-1} e^{-\tau X} \\
        & \rightarrow \frac{1}{(n-1)!}\int^{1/\Lambda^2_{IR}}_{1/\Lambda^2_{UV}} d \tau \tau^{n-1} e^{-\tau X},
    \end{split}
\end{equation}
where $X$ represents a product of propagators with Feynman parameterization. The ultraviolet cutoff, $\Lambda_{UV}$, and infrared cutoff, $\Lambda_{IR}$, have been determined in Ref.~\cite{Cloet:2014rja}. The cutoff parameters in the proper-time regularization was taken as used in Ref.~\cite{Carrillo-Serrano:2015uca}, $\Lambda_{IR} = 0.240 \,\text{GeV} $ and $\Lambda_{UV} = 0.645 \,\text{GeV}$.

\section{Numerical Results}\label{section3}

\subsection{Model parameters}

\begin{figure}[h]
	\centering
	\includegraphics[width=7cm]{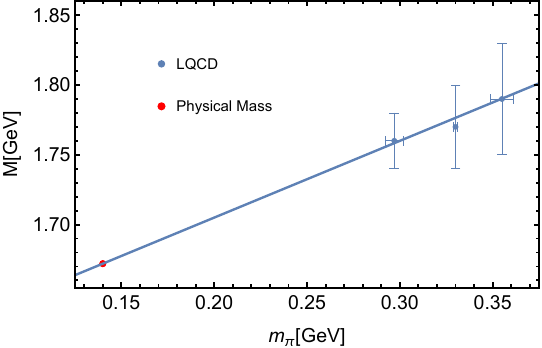}
	\caption{\small{The non-physical masses $M$ of $\Omega^-$ baryon as a function of $\pi$ mass $m_\pi$ calculated by lattice QCD~\cite{Alexandrou:2010jv}. Red dot located in the lower left corner represents the physical masses of the $\Omega^-$ and pion.}}
	\label{figure-non-physical-masses}
\end{figure}

There are three model parameters, $c_{2}$, $c_{3}$ and $m_R$, which need to adjust to fit the EMFFs obtained from the lattice QCD calculations. It is worthy to mention that the lattice QCD calculations~\cite{Alexandrou:2010jv} are obtained for three pion masses in the range of about 300 to 350 MeV, namely $m_\pi = 297$ MeV, $m_\pi = 330$ MeV, and $m_\pi = 350$ MeV, which are not at the physical masses of included particles. In this work, from the non-physical masses $M$ of $\Omega^-$ baryon as a function of $\pi$ mass $m_\pi$ calculated by lattice QCD, we take a linear function $M = 0.56 \, m_\pi + 1.5936 \,\text{GeV}$ to fit the relation between $M$ and $m_\pi$.
The fitted results are shown in Fig.~\ref{figure-non-physical-masses}. One can see that we can get a fairly good fit.
The other masses included in this work are also adjusted accordingly.
We take the change rate of the strange quark is one third of the $\Omega^-$ mass and the diquark two thirds, and, the light quark and Kaon roughly take the same change rate with the strange quark due to the similar masses. 
After calculation, $Z$ in Eq.~\eqref{self-energy-Z} and $Z_K$ in Eq.~\eqref{tauk} corresponding to the physical masses respectively are $Z=0.907$ and $Z_K=20.28$.

Besides, from fitting to the lattice results with different masses of $\pi$, we get the model parameters, $c_2=0.13 \,\text{GeV}^{-1}$, $c_3=0.05 \,\text{GeV}^{-2}$ and $m_R = M - 0.1 \, \text{GeV}$ in all the situations. At the physical mass point of pion, we get the physical $\Omega^-$ mass $M = 1.672 \,\text{GeV}$, and the corresponding strange quark mass $m_s = 0.6 \,\text{GeV}$ and diquark mass $m_D = 1.15 \, \text{GeV}$ are consistent with our previous work~\cite{Fu:2023ijy}. Besides, we take the Kaon mass $m_K = 0.494 \,\text{GeV}$ and light quark mass $m_l = 0.4 \,\text{GeV}$. All these above mass parameters used in this work are listed in Tab.~\ref{table-parameters}.

\begin{table}[htbp]
	\renewcommand\arraystretch{1.3}
	\centering
	 \caption{\small{Mass parameters used in this work. The unit of these masses is in \text{GeV}.}} 
 \begin{tabular}{ c c c c c c c}
		 \toprule
		 \toprule
		  $m_\pi$ & $M$	&  $m_s$ & $m_D$ & $m_l$	& $m_K$\\
		\midrule
		 0.350	& 1.792	& 0.640	& 1.230 & 0.440	& 0.534 \\
		 0.330	& 1.778 & 0.636	& 1.221 & 0.436 & 0.530 \\
		 0.297	& 1.760 & 0.629	& 1.209 & 0.429 & 0.523	\\
		 0.140	& 1.672 & 0.600	& 1.150 & 0.400 & 0.494	\\
		\bottomrule
		\bottomrule
	 \end{tabular}
	\label{table-parameters}  
\end{table}

\subsection{ The $\Omega^-$ EMFFs in the spacelike region} \label{section-results-SL}

\begin{figure}[htbp]
	\centering
	\includegraphics[width=7cm]{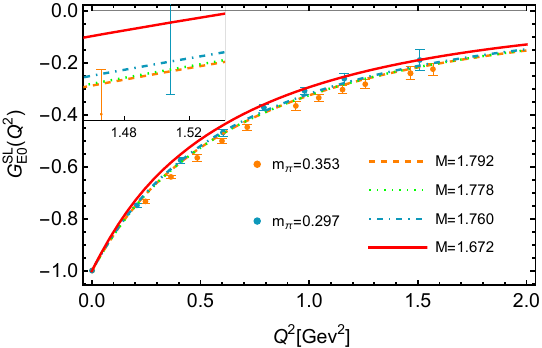} \quad
	\includegraphics[width=7cm]{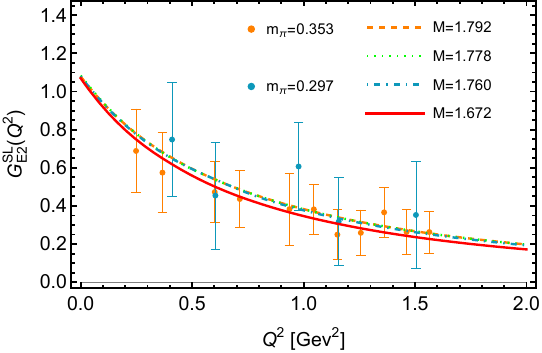} \\
	\includegraphics[width=7cm]{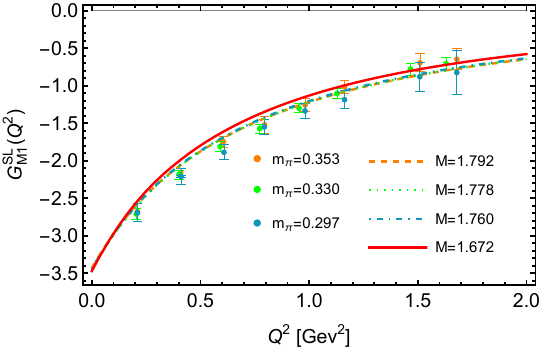} \quad
	\includegraphics[width=7cm]{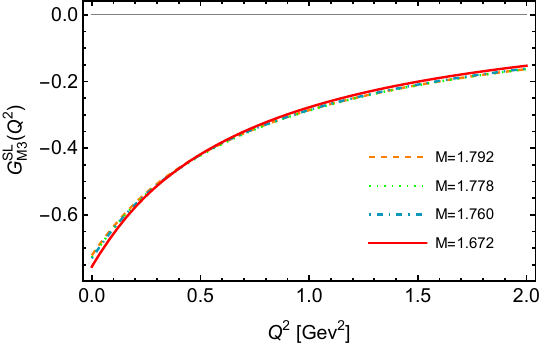}
	\caption{\small{The obtained EMFFs in the spacelike region for the $\Omega^-$ baryon corresponding to the different valuse of $m_\pi$. The mass dimension GeV of $m_\pi$ and $M$ is omitted in the legends.}}
	\label{figure-EMFFs}
\end{figure}

With the model parameters shown in Tab.~\ref{table-parameters}, all the EMFFs for the $\Omega^-$ baryon in the spacelike region can be easily obtained and we show them in Fig.~\ref{figure-EMFFs}, where the orange dashed, green dotted, and blue dot dashed lines respectively represent the EMFFs corresponded to pion mass, $0.353 \,\text{GeV}$, $0.330 \,\text{GeV}$, and $0.297 \,\text{GeV}$ (or $\Omega^-$ mass, $1.792 \,\text{GeV}$, $1.778 \,\text{GeV}$ and $1.760 \,\text{GeV}$). It is found that the lattice QCD calculations can be well reproduced. And one can conclude that the variation tendency is consistent as the mass approaches the physical mass by observing the enlarged view in the top left corner of the first panel $G_{E0}^{SL}$. Comparing the results obtained with non-physical and physical masses, the real radii of charge and magnetic moment are larger than the lattice results. This can be clearly seen from the inset of Fig.~\ref{figure-EMFFs} for the numerical results of $G^{\rm SL}_{E0}(Q^2)$ at the range of $1.45 < Q^2 < 1.55$ ${\rm GeV}^2$.

\begin{figure*}[htbp]
	\centering
\includegraphics[width=18cm]{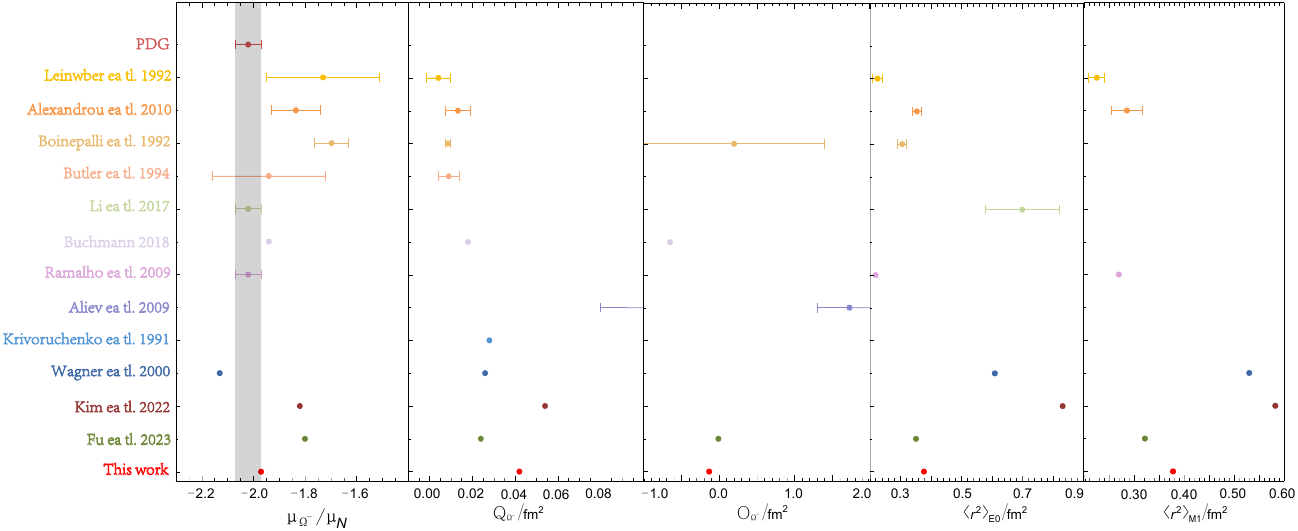}
\caption{The magnetic moment, electric-quadrupole moment, magnetic-octupole moment, electric charge radius and magnetic radius in comparison with those from PDG~\cite{ParticleDataGroup:2024cfk}, LQCD~\cite{Leinweber:1992hy,Alexandrou:2010jv,Boinepalli:2009sq}, $\chi$PT~\cite{Butler:1993ej,Li:2016ezv}, $1/N_c$ expansion~\cite{Luty:1994ub,Buchmann:2018nmu}, relativistic quark model~\cite{Ramalho:2009gk}, non-relativistic quark model~\cite{Krivoruchenko:1991pm}, QCD sum rules~\cite{Lee:1997jk,Aliev:2009pd}, $\chi$ quark model~\cite{Wagner:2000ii}, and chiral quark-soliton model~\cite{Kim:2019gka,Jun:2021bwx}.} \label{FigureEMFFs}
\end{figure*}

In the forward limit, $Q^2=0$, we can get the magnetic moment $\mu_{\Omega^-} = G^{SL}_{M1}(0) M_N \mu_N/M$, electric-quadrupole moment $\mathcal{Q}_{\Omega^-} = G^{SL}_{E2}(0) |e|/M^2$, and magnetic-octupole moment $\mathcal{O}_{\Omega^-} = G^{SL}_{M3}(0) M^3_N \mathcal{O}_N /M^3$ (note that $G^{SL}_{M3}(0)$ has the coefficient difference from Ref.~\cite{Korner:1976hv} as mentioned above). The physical quantities of $\mu_N$ and $\mathcal{O}_N$ with the subscript $N$ stand for the corresponding nuclear properties and $M_N$ being the proton mass. With the contribution of the Kaon meson cloud, the magnetic moment, $\mu_{\Omega^-}=-1.97 \mu_N$, is consistent with the experimental measurements as quoted in PDG~\cite{ParticleDataGroup:2024cfk}. A comparison of our numerical results with those obtained with other models is shown in Fig.~\ref{FigureEMFFs}. One can see that for the properties of $\Omega^-$ baryon, our numerical results are lied between those of other model calculations. It is expected that future experimental measurements can be used to test these model calculations.

It was known that the meson cloud makes the charge distribution in the quark more dispersed and further causes the magnetic moment and radii to increase. Considering Eq.~\eqref{dressed-quark-vertex}, the dressed quark with the meson cloud compared with the point-like quark has an extra term $i \sigma^{\mu q}/(2 M)$, which plays a significant role in the magnetic form factors.

In this work, the electromagnetic radii of $\Omega^-$ are obtained as,
\begin{eqnarray}
\langle r^2\rangle_{E0} &=& 0.378 ~ \text{fm}^2, \\
\langle r^2\rangle_{M1} &=& 0.378 ~ \text{fm}^2.
\end{eqnarray}
It is found that the above electric and magnetic radii are same and the magnetic radius has changed much compared with the value obtained in Ref.~\cite{Fu:2023ijy}, where the affect of the meson cloud was not taken into account. On the other hand, the changes of the electric radius is small. This can be explained by the fact that the meson cloud has a greater impact on the magnetic form factor than on the electric form factor.

\begin{figure*}
	\centering
	\includegraphics[width=7.5cm]{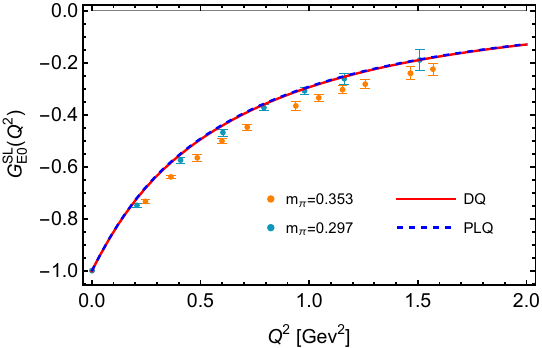} \quad
	\includegraphics[width=7.5cm]{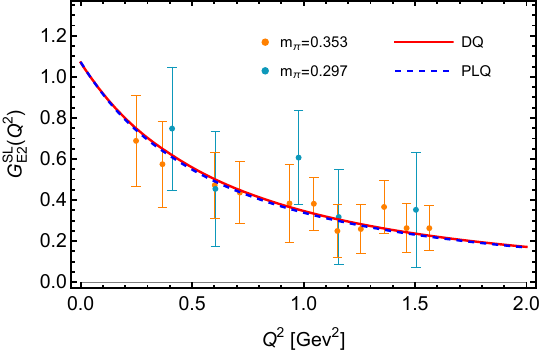} \\
	\includegraphics[width=7.5cm]{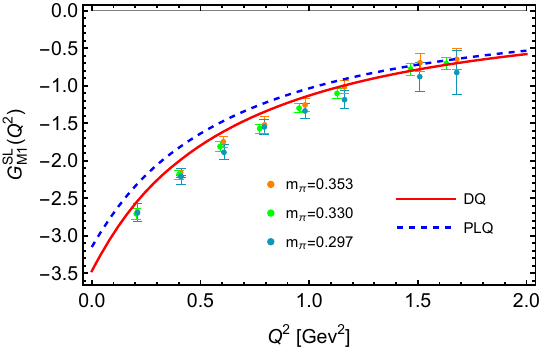} \quad
	\includegraphics[width=7.5cm]{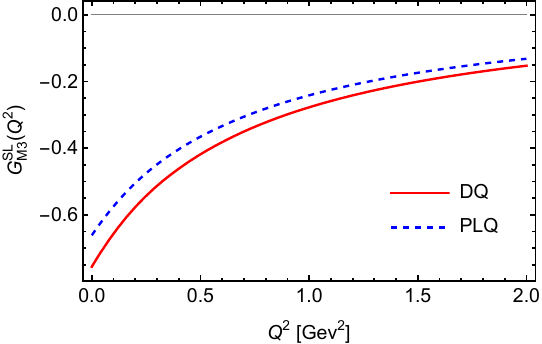}
\caption{\label{figure-EMFFs-PL}\small{The EMFFs comparison between the point-like quark and dressed quark under the same parameters. And the mass dimension GeV of $m_\pi$ is omitted in the legends. DQ and PLQ in the legends respectively represent the dressed quark with the Kaon cloud and the point-like quark, and these two superscript labels are applicable to the full text.}}
\end{figure*}

To see more clearly the effects of the Kaon meson cloud, we show the $\Omega^-$ EMFFs obtained with and without Kaon meson cloud in Fig.~\ref{figure-EMFFs-PL} using the physical masses of included particles. Again, one can see that the Kaon meson cloud has minimal influence on the electric form factors and can be neglected, but it gives significant contribution on the magnetic form factors. For the case of $G^{\rm SL}_{M1}(Q^2)$, the theoretical results with the contribution of Kaon meson cloud are more close to the lattice QCD calculations, comparing with the numerical results without considering the effects of Kaon meson cloud. This may indicate that the meson cloud contributions should be included. Nevertheless, the charge distribution of the zero charge particle like the neutron~\cite{Wang:2024abv} extremely dependant on the meson cloud due to the larger relative change of $F_1$ in Eq.~\eqref{dressed-quark-vertex}. Further model calculations are needed to clarify this issue.

\begin{figure}
	\centering
	\includegraphics[width=8cm]{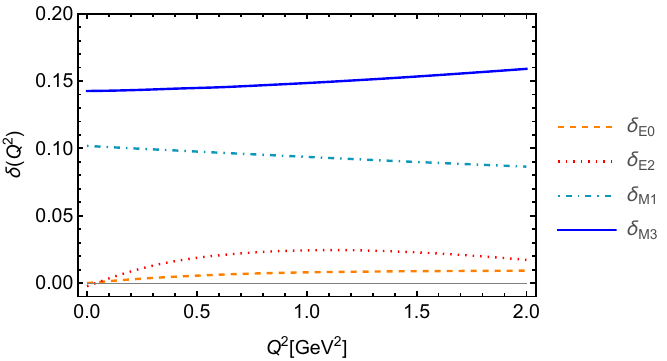}
\caption{\label{figure-delta-EMFFs-SL}\small{The difference of the $\Omega^-$ EMFFs in the spacelike region obtained with and without the contributions from the Kaon meson cloud.}}
\end{figure}

Next, in order to intuitively illustrate the effect from the Kaon meson cloud, we define
\begin{equation}
    \delta_i(Q^2) = \frac{G^{SL,{\rm DQ}}_i(Q^2)- G^{SL,{\rm PLQ}}_i(Q^2)}{G^{SL,{\rm PLQ}}_i(Q^2)},
\end{equation}
which stands for the relative differences of $\Omega^-$ EMFFs between results obtained with and without the Kaon meson cloud. The $i$ labels the different form factors. The numerical results of $ \delta_i(Q^2) $ are shown in Fig.~\ref{figure-delta-EMFFs-SL}, from where it is found that the Kaon meson cloud give larger contributions to the magnetic form factors than the electric form factors and large impact to the higher-order terms. Indeed, one can approximately extract the change rates of the different form factors, $\delta_{E0}\approx 1 \%$, $\delta_{E2} \approx 2 \%$, $\delta_{M1} \approx 10 \%$ and $\delta_{M3} \approx 15\%$. Moreover, the rates of difference because of the Kaon meson cloud remain almost unchanged as the $Q^2$ increases. In other words, the meson cloud plays an important role in all the energy region. 

\subsection{The $\Omega^-$ EMFFs in the timelike region}

\begin{figure}
	\centering
	\includegraphics[width=7.5cm]{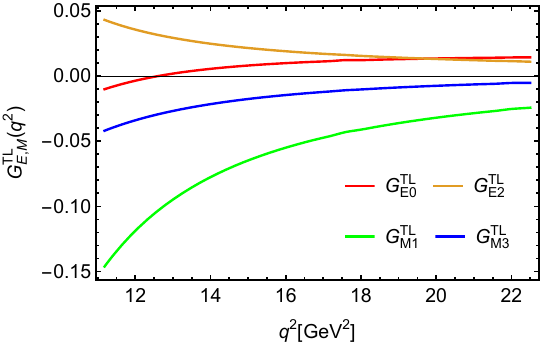}
	\caption{\small{The theoretical results of the EMFFs for $\Omega^-$ baryon as a function of $q^2$ at the timelike region.}}
	\label{figure-TLEMFFs}
\end{figure}

The asymptotic relations of Eq.~\eqref{SL_to_TL} allow us to extend the EMFFs from the spacelike to the timelike region. With all the obtained model parameters in the spacelike region, we can calculate the electromagnetic form factors of $\Omega^-$ in the timeklike region. The theoretical predictions of the EMFFs for $\Omega^-$ baryon at the timelike region~\footnote{In general, electromagnetic form factors in the timeklike region are complex. However, only the real timelike form factors can be obtained in this work due to the real spacelike form factors.}
are shown in Fig.~\ref{figure-TLEMFFs} as a function of $q^2$. The electric form factors $G^{TL}_{E0}(q^2)$ and $G^{TL}_{E2}(q^2)$ are presented by red and yellow curves, respectively. The magnetic form factors $G^{TL}_{M1}(q^2)$ and $G^{TL}_{M3}(q^2)$ are plotted by the green and blue lines, respectively. In the whole energy region considered here, $G^{TL}_{E2}(q^2)$ is larger then zero, and $G^{TL}_{M1}(q^2)$ and $G^{TL}_{M3}(q^2)$ are less than zero. Yet, the form factor $G^{TL}_{E0}(q^2)$ is zero at $q^2 \sim 12.2 \,\text{GeV}^2$. Note that at the limit $q^2 \sim \infty$, all the four electromagnetic form factors are going to zero.

\begin{figure}
	\centering
	\includegraphics[width=8cm]{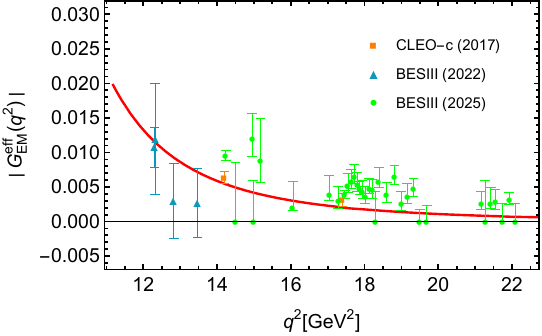} 
	\caption{\small{The obtained $\Omega^-$ effective form factor $|G^{eff}_{EM}(q^2)|$ with the dressed quark under the same parameters comparing with the CLEO-c data (orange dots~\cite{Dobbs:2017hyd}) and BESIII data (blue dots~\cite{BESIII:2022kzc} and green dots~\cite{BESIII:2025azv}). }}
	\label{figure-EF-EMFFs}
\end{figure}

Then, one can also obtain the effective form factor $|G^{eff}_{EM}(q^2)|$ of the $\Omega^-$ baryon as defined in Eq.~\eqref{Effective_EMFFs}. The theoretical results of the effective form factor are shown in Fig.~\ref{figure-EF-EMFFs}, where the numerical results are obtained with the Kaon meson cloud as the red solid line. The experimental data are taken from Refs.~\cite{Dobbs:2017hyd,BESIII:2022kzc,BESIII:2025azv} and they were attracted from the total cross sections of $e^+ e^- \rightarrow \Omega^- \bar{\Omega}^+$ reaction. One can see that the theoretical result can describe the experimental measurements within uncertainties.
Therefore, more precise experiments, especially near the threshold, are needed to test the accuracy of the model.

Next, we study the polarization of $\Omega^-$ baryon in the $e^+ e^- \rightarrow \Omega^- \bar{\Omega}^+$ reaction. In the helicity framework, the production of $\Omega^- \bar{\Omega}^+$ in $e^+ e^-$ annihilation process is analyzed in the center-of-mass frame of $e^+ e^-$. To analyze the polarization of the single-tag $\Omega^-$, one can sum over the spin states of $\bar{\Omega}^+$ and the spin states of the initial $e^+$ and $ e^-$. And the differential cross section of $e^+ e^- \rightarrow \Omega^- \bar{\Omega}^+$ reaction including the nonzero polarization components is given as follows~\cite{Zhang:2023box}:

\begin{eqnarray}\label{differential_cross_section}
\frac{d \sigma}{d cos \theta} &=& \frac{d \sigma^{un}}{d cos \theta} + \frac{d \sigma_{LL}}{d cos \theta}S_{LL} +  \nonumber \\ 
&&  \frac{d \sigma_{LT}^{x}}{d cos \theta}S_{LT}^{x} + \frac{d \sigma_{TT}^{xx}}{d cos \theta}S_{TT}^{xx},
\end{eqnarray}
with
\begin{widetext}
\begin{subequations}\label{po_cross_section}
	\begin{equation}
		\frac{d \sigma_{LL}}{d cos \theta} = \frac{\pi \alpha^2 \beta C}{2 q^2} \left[\frac{8}{3}\text{Re}\left[G^{TL}_{E0} G^{TL *}_{E2}\right]\text{sin}^2\theta - \left(\frac{4}{9}\left| G^{TL}_{M1}\right|^2 +\frac{16}{25} \tau^2 \left|G^{TL}_{M3}\right|^2 - \frac{16}{15} \tau \text{Re}\left[G^{TL}_{M1} G^{TL *}_{M3}\right] \right)(1+\text{cos}^2\theta)\right],
	\end{equation}
	\begin{equation}
		\frac{d \sigma_{LT}^x}{d cos \theta} = \frac{\pi \alpha^2 \beta C}{2 q^2} \frac{1}{15 \sqrt{\tau}}\left(\left|3 G^{TL}_{E0} + 2 \tau G^{TL}_{E2} \right| - \left|3 G^{TL}_{E0} -2 \tau G^{TL}_{E2} \right|\right)\left|5 G^{TL}_{M1} +4 \tau^2 G^{TL}_{M3} \right|\text{sin}(2 \theta),
	\end{equation}
	\begin{equation}
		\frac{d \sigma_{TT}^{xx}}{d cos \theta} = \frac{2 \pi \alpha^2 \beta C}{3 q^2} \left| G^{TL}_{M1} - \frac{6}{5} \tau G^{TL}_{M3} \right| \left| G^{TL}_{M1} + \frac{4}{5} \tau G^{TL}_{M3} \right| \text{sin}^2\theta.
	\end{equation}
\end{subequations}
\end{widetext}
where the $S_{LL}$, $S_{LT}^x$ and $S_{TT}^x$ represent the rank-2 spin tensor of the outgoing $\Omega^-$. Summing over all the polarization states of the final $\Omega^-$, the differential cross section in Eq.~\eqref{differential_cross_section} will be simplified to the unpolarized differential cross section. Moreover, three additional differential cross sections have been ignored because the EMFFs in the timelike region are also real in this work.
In the following, all the results are calculated and discussed on the basis of the real EMFFs and unpolarized initial states of $e^+$ and $e^-$.

\begin{figure}[htbp]
	\centering
	\includegraphics[width=7cm]{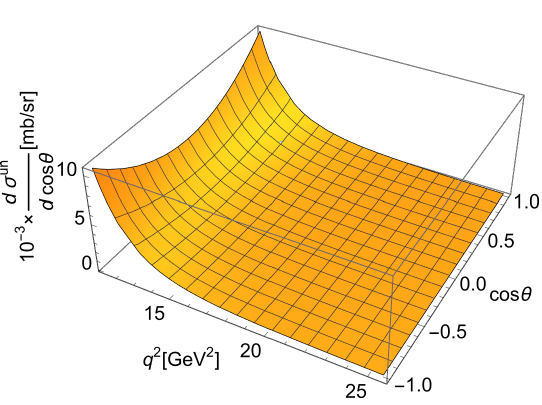} \quad
	\includegraphics[width=7cm]{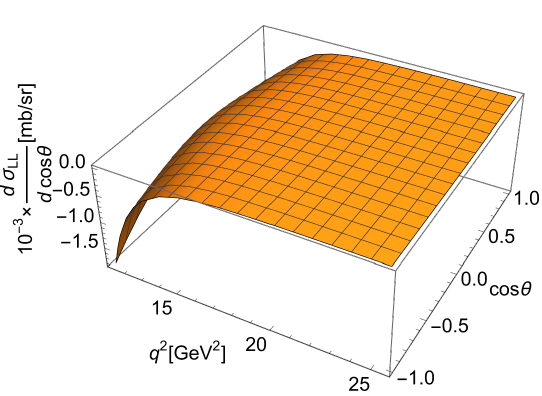}\\
	\includegraphics[width=7cm]{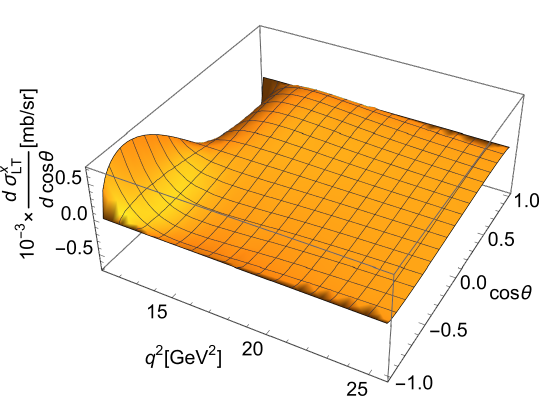} \quad
	\includegraphics[width=7cm]{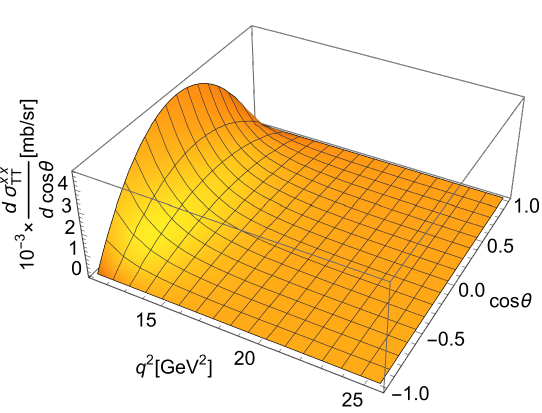}\\
	\caption{\label{fig:cross-section}\small{Double differential cross sections with different polarized final $\Omega^-$ in $e^+ e^- \rightarrow \Omega^- \bar{\Omega}^+$ as functions of $q^2$ and scattering angle ${\rm cos}\theta$.}}
\end{figure}

With the model parameters, we can get the unpolarized and nonzero double differential cross sections of $e^+ e^- \to \Omega^- \bar{\Omega}^+$ reaction with different polarized $\Omega^-$ baryon as a function of $q^2$ and the scattering angle ${\rm cos}\theta$. The numerical results are shown in Fig.~\ref{fig:cross-section}. In the parity-conserving $e^+ e^- \rightarrow \Omega^- \bar{\Omega}^+$ process, mirror transformation about the plane perpendicular to the direction of the $\Omega^-$, equivalent to $\theta \to \pi - \theta$, contribute a minus sign to the transverse polarization. Therefore, the differential cross section $\frac{d \sigma^{un}}{d \text{cos} \theta}$, $\frac{d \sigma_{LL}}{d \text{cos} \theta}$ and $\frac{d \sigma_{TT}^{xx}}{d \text{cos} \theta}$ are $\text{cos}\theta$-even, but $\frac{d \sigma_{LT}^{x}}{d \text{cos} \theta}$ is $\text{cos}\theta$-odd.

When the direction of the $\Omega^-$ baryon coincides with the electron beam, the motion direction of all the included particles are in the same line, then the differential cross sections of $e^+ e^- \to \Omega^- \bar{\Omega}^+$ should be transverse independent. Furthermore, the polarized differential cross sections with transverse part will be zero at $\theta = 0$ and $\pi$. These features can be seen in Fig.~\ref{fig:cross-section}.

\begin{figure*}
	\centering
	\begin{minipage}[]{0.44\textwidth}
		\subfigure[]{\includegraphics[width=8 cm]{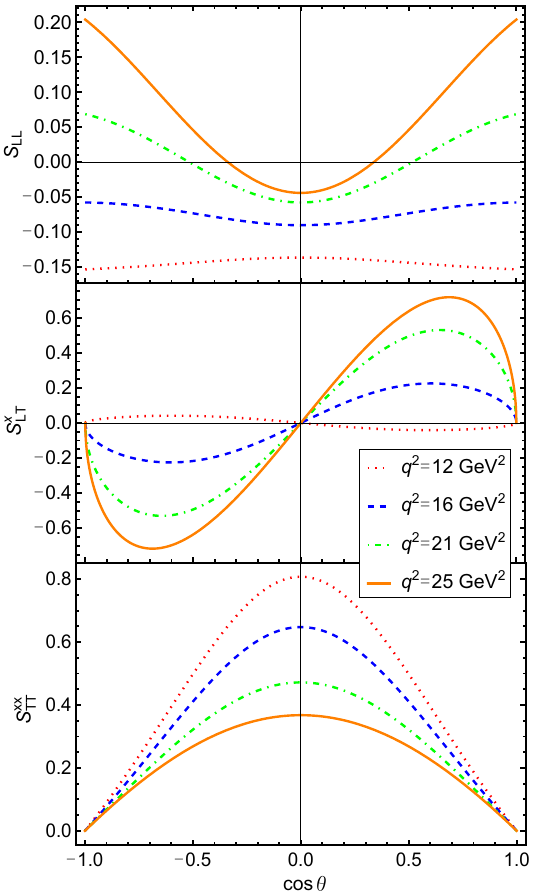}}
	\end{minipage}
	\begin{minipage}[]{0.44\textwidth}
		\subfigure[]{\includegraphics[width=5 cm]{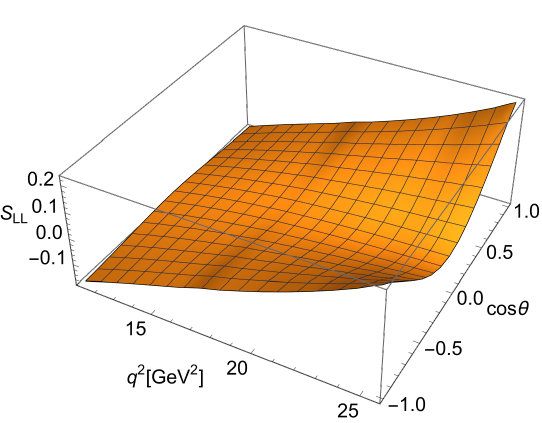}}\\
		\subfigure[]{\includegraphics[width=5 cm]{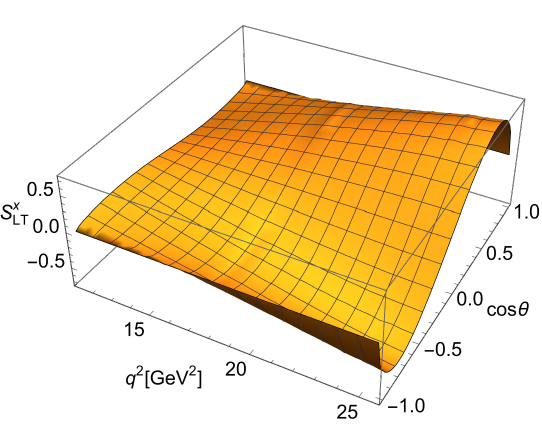}}\\
		\subfigure[]{\includegraphics[width=5 cm]{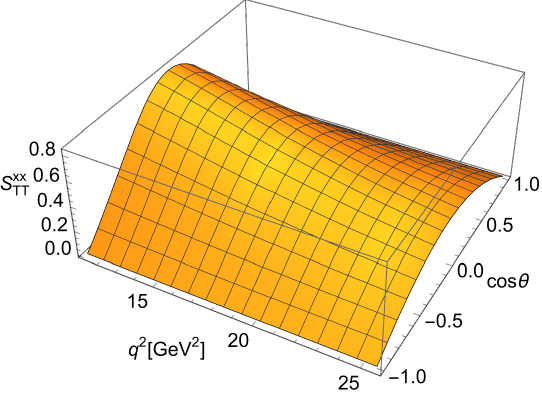}}
	\end{minipage}
	\caption{\label{fig:spin_components}\small{Spin components in the single-tag $\Omega^-$ process.}}
\end{figure*}

Based on the obtained differential cross sections, one can then obtain the spin components of the final $\Omega^-$ baryon. For doing this, we define the spin components as
\begin{eqnarray}
S_{X} = \frac{d \sigma_{X}}{d \text{cos} \theta}/\frac{d \sigma^{un}}{d \text{cos} \theta}, 
\end{eqnarray}
where $X$ stands for the different polarization. For convenience and clarity, $S_{X}$ represents the polarization direction in Eq.~\eqref{differential_cross_section} and spin component in the numerical results.
The spin components of the final $\Omega^-$ are respectively shown as the 2D functions of $\text{cos} \theta$ with different energy and the 3D functions of $\text{cos}\theta$ and $q^2$ in Fig.~\ref{fig:spin_components}.
From Fig.~\ref{fig:spin_components} (a) (c) and (d), one can also obviously find that the final state is transverse unpolarized when the outgoing direction coincide with the electron beam. Moreover, all the results satisfy the boundary constraints of Fig.~2 in Ref.~\cite{Zhang:2023box}.
Figure~\ref{fig:spin_components} (a) give the slices of the spin components with different energy, and they imply that the longitudinal part is primary at high energy region and the transverse part is significant at low energy region. With the increase of energy, $S_{LL}$ becomes large and $S^{xx}_{TT}$ is trending to zero but $S^x_{LT}$ is confusing because it has both the longitudinal and transverse parts. Furthermore, the results elaborates that the spin components will concentrate on the direction of the electron beam as energy increasing, and then one can approximately presume that the polarization of the final single-tag $\Omega^-$ only exist at the initial particle direction in the high energy limit.

\section{Summary and Conclusion}\label{section4}

Based on the quark-diquark approach with the meson cloud effect, we study the EMFFs of $\Omega^-$ hyperon both in the spacelike and timelike regions. Considering the $\Omega^-$ with $I(J^P)=0(\frac{3}{2}^+)$ as the combination of a strange quark and an axial-vector diquark with two strange quarks can simplify the three-body system to the two-body system. The diquark is a two-body subsystem and it was computed separately. To fit the lattice results precisely, the Kaon meson cloud is also taken into account and other mesons are omitted because Kaon is the lightest meson with strange quark and plays a primary role. The Kaon meson cloud introduces the term $\frac{i \sigma^{\mu q}}{2M}$, which is absent in the description of a point-like quark. Furthermore, the term $\frac{i \sigma^{\mu q}}{2M}$ mainly contributes to the magnetic-dipole and -octupole form factors by the definition and this feature has been verified in the numerical results.

The model free parameters were adjusted by fitting them to the lattice calculations for the EMFFs of the $\Omega^-$ baryon in the spacelike region. With the fitted parameters and the physical masses of included particles, the genuine EMFFs in the spacelike region are obtained. The magnetic moment from this approach is more close to the experimental value and the radii of charge and magnetic are larger than the lattice and our previous work.

Employing the asymptotic relations, we extend the EMFFs in the spacelike region to the timelike region with the shift of $2 M^2$. Then the principal results, the effective form factor of $\Omega^-$ baryon, was obtained and is acceptable by comparing with the experimental data from CLEO and BESIII. By comparing with lattice QCD calculations and the experimental measurements, our results imply that the quark-diquark approach with the Kaon meson cloud is reliable when describing the electromagnetic interaction in both the spacelike and timelike regions. An important feature of the Kaon meson cloud is that the relative influence from it keeps almost unchanged as the energy changed. Therefore, the change rates can help us to estimate other related physical quantities.

We also examined the polarized differential cross sections for \( e^+ e^- \rightarrow \Omega^- \bar{\Omega}^+ \) in the helicity formalism, focusing on the single-tag \(\Omega^-\) polarization with unpolarized initial states of $e^+$ and $e^-$ and without considering the phase information of the EMFFs. The differential cross section is split into unpolarized and polarized components, linked to the \(\Omega^-\) spin tensor. The \(S_{LL}\) and \(S_{TT}^{xx}\) terms are \(\cos\theta\)-even, while \(S_{LT}^x\) is \(\cos\theta\)-odd, with transverse polarization vanishing at \(\theta = 0, \pi\). Numerical results reveal that longitudinal polarization dominates at high energies, while transverse contributions are significant at low energies, and the existence range converges toward alignment with the electron beam direction in the high-energy limit. It is expected that these theoretical predictions can be tested by future experiments that can be carried out by the Belle II, LHCb, and BESIII Collaborations in the near future, as well as by the next generation facilities~\cite{KLF:2020gai,Aoki:2021cqa}.\\

\section{Acknowledgements}

One of us (D.Y. Fu) is grateful to Zhe Zhang for constructive discussions. This work is partly supported by the National Key R\&D Program of China under Grant No. 2023YFA1606703, and by the National Natural Science Foundation of China under Grants Nos. 12375142, 12435007, 12361141819 and 12447121. It is also supported by the Gansu Province Postdoctor Foundation.

\bibliographystyle{unsrt}
\bibliography{refs}

\end{document}